\documentclass[cits,hyper]{JINST}

\usepackage{amsmath}

\newcommand{\di}{\ensuremath{\,\mathrm{d}}}         
\newcommand{\eff}{\ensuremath{\varepsilon}}         
\newcommand{\Bi}{\ensuremath{\mathrm{Bi}}}          
\newcommand{\Be}{\ensuremath{\mathrm{Be}}}          
\newcommand{\vect}[1]{\ensuremath{\boldsymbol{#1}}} 
\newcommand{\Pt}{\ensuremath{\vect{p}_\text{T}}}  

\newcommand{\doi}[1]{\href{http://dx.doi.org/#1}{doi: #1}}
\newcommand{\arxiv}[1]{\href{http://arXiv.org/abs/#1}{arXiv: #1}}
\newcommand{\url}[1]{\href{#1}{#1}}

\title{Estimating the selection efficiency}

\author{Diego Casadei\\
  New York University, 
  4 Washington Place, NY-10003 New York, USA
  \\
  \email{diego.casadei@cern.ch}
}

\abstract{%
  The measurement of the efficiency of an event selection is always an
  important part of the analysis of experimental data.  The
  statistical techniques which are needed to determine the efficiency
  and its uncertainty are reviewed.  Frequentist and Bayesian
  approaches are illustrated, and the problem of choosing a meaningful
  prior is explicitly addressed.  Several practical use cases are
  considered, from the problem of combining different samples to
  complex situations in which non-unit weights or non-independent
  selections have been used.
}


\begin{document}




\section{Introduction}

 There are several cases in high-energy physics (HEP) in which one has
 to measure a selection efficiency, for example when dealing with the
 trigger or offline event selection with the aim of measuring a cross
 section.  The selection efficiency is the conditional probability
 that any single event passes the selection, given all other
 conditions (type and energy of the collisions, detector
 configuration, and possibly a preselection).  A good estimator of the
 selection efficiency is the measured success frequency, that is the
 ratio between the number of surviving events and the size of the
 initial sample, which for larger and larger samples converges in
 probability to the selection efficiency (Bernoulli's theorem).
 Different ways of summarizing the uncertainty on the true efficiency
 exist, as explained below, within the frameworks of the traditional
 (frequentist) and of the Bayesian approaches.

 A short comparison between the two approaches is provided in
 \cite[chapter~32]{pdb08} and a more complete comparison can be found
 in \cite[appendix~B]{BayesianTheory1994}.  The frequentist approach,
 often preferred in HEP data analysis, is fully reviewed by
 \cite{cousins2010}, including the open issues related to the
 possibility of choosing among different solutions for the same
 problem.  Although in the Bayesian framework there is only one way of
 finding the solution, freedom remains in the choice of the prior, as
 it will be discussed below.

 The most important difference among the two approaches is the kind of
 question they address.  In the classical approach, the typical answer
 is in terms of the probability of obtaining some result, given the
 model.  On the other hand, the Bayesian approach answers questions
 about the probability that some hypothesis is true, given the
 observed data.  Hence, once the question is formulated in terms of
 either the probability of the data given the model or the probability
 of the hypothesis given the observation, one knows which approach is
 to be selected.
 In our case, the statistical inference deals with the
 unknown value of the parameter of interest, the selection efficiency.
 In the Bayesian approach, this is reduced to a problem of
 probability, because the framework provides the solution in terms of
 the probability distribution of the unknown parameter (interpreted as
 a description of the experimenter's degree of belief on the possible
 values) given the observation.  On the other hand, this probability
 distribution is not even defined in the frequentist approach, in
 which one considers instead (hypothetical) identical repetitions of the
 same experiment and looks at the fraction of times in which the
 result is expected to be compatible with the (unknown) true value.


 In most practical problems, the experimenter needs either a single
 value (the best estimate of the efficiency), to be used in the
 computation of the cross section, or a ``reasonable'' interval which
 is supposed to contain the true (unknown) value with high
 probability, to be used when reporting the result.  In the
 frequentist approach, such interval is reported in the form of a
 \emph{confidence interval} associated to its coverage, for example a
 68.3\% or a 95\% confidence interval.  This is interpreted in terms
 of identical repetitions: a coverage of $q$ means that, in the limit
 of a very high number of replications, the fraction of intervals
 (constructed in the same way) which contain the true value is $q$.
 It is important to note that this does \emph{not} mean --- as it is
 sometimes implicitly assumed --- that the true value has $q$
 probability of being contained by the reported interval.  In
 contrast, this is exactly the interpretation of a $q$-credible
 interval obtained in the Bayesian framework.  When the sample size is
 very large, the Bayesian $q$-credible intervals will have also
 coverage of $q$, although the rapidity of this convergence is
 different for different choices of the prior.

 Our model can be considered a (possibly infinite) sequence $r_1, r_2,
 \ldots \in \{0,1\}$ of Bernoulli's random variables which either
 posses or not a given property.  The selection process consists in
 discarding those ``events'' which do not have that property, and the
 selection efficiency corresponds to the long run relative frequency
 $k/n$ where $k = \sum_{i=1}^{n} r_n$ is the total number of surviving
 events out of the initial $n$ events.  This is the result of the
 Bernoulli's theorem, also known as the ``weak law of large numbers'':
 the relative frequency will converge in probability to the true
 efficiency in the limit of an infinite number $n$ of
 measurements.\footnote{Convergence in probability means that
 ``unusual'' outcomes become less and less likely as the sequence
 $\{r_n\}$ of random variables progresses, and it is weaker than
 mathematical convergence, for which there exist some $n_0$ such that
 it never happens that, for $n>n_0$, the maximum allowed distance
 from the limit is exceeded.}

 In the usual case in which the total number $n$ of initial events is
 fixed independently from $k$ (for example, when it depends on the
 accelerator or detector live time), the probability of selecting $k$
 events, given the sample size $n$ and the selection efficiency \eff,
 is given by the \emph{binomial distribution}
\begin{equation}\label{eq-binomial}
  P(k| \eff, n) = \Bi(k| \eff, n)
                \equiv \binom{n}{k} \eff^{k} (1-\eff)^{n-k}
  \; .
\end{equation}
 The likelihood function of the model $L(\eff;n,k)$ is also given by
 (\ref{eq-binomial}), in which $\Bi(k|\eff,n)$ is interpreted as a
 function 
 of the unknown efficiency
 \eff\ with data fixed at the observed counts $k$ given~$n$.

 In the (less common) cases in which one continues performing the
 selection until $r\ge1$ events are collected, the probability model
 is the \emph{negative binomial} distribution:
\begin{equation}\label{eq-negative-binomial}
  P(n| \eff, r) = \text{Nb}(n| \eff, r)
                \equiv \binom{n-1}{r-1} \eff^{r} (1-\eff)^{n-r}
\end{equation}
 with $n=r, r+1, r+2, \ldots$

 The frequentist approach is summarized is
 section~\ref{sec-freq-approach} whereas the Bayesian approach is
 explained in section~\ref{sec-Bayes-approach}.  Finally, several use
 cases are considered in detail in section~\ref{sec-patologic}.

\section{The classical approach}\label{sec-freq-approach}

 Most often, we deal with the binomial model and are interested into
 the selection efficiency as a function of some measured quantity $x$,
 which can be a scalar (e.g.~the missing transverse momentum in the
 event) or a multidimensional parameter (e.g.~the magnitude of the
 transverse momentum and the two angles defining the direction of a
 reconstructed electron).  In practice, we fill histograms of $x$
 before and after the selection $S$ and compare the entries $k_i$
 (surviving the selection $A$) and $n_i$ (initial counts) in each bin
 $i$.  A bin-wise ratio of the histograms filled after and before the
 selection will result into the histogram of the success frequencies
 $f_i = k_i/n_i$, which, by virtue of Bernoulli's theorem, can be
 taken as the estimates of the unknown efficiencies $\eff_i$ (the
 histograms \emph{must} be filled with unit weights for this to be
 true\footnote{Integer weights are allowed to the extent in which one
   event with weight $w$ is just a short-hand notation for $w$ events
   which \emph{all} either pass or fail the selection.}):
 \begin{equation*}
   \eff_i \equiv 
   \int_{\text{bin $i$}} \eff(x;A) \di x
   \; \approx \; f_i(A)
   \equiv \frac{k_i(A)}{n_i}
\end{equation*}
 where we use the symbol $\approx$ to mean ``is estimated by''.  In
 the following, we will omit the bin index ($i$) and the selection
 ($A$) for simplicity.

 The frequency $f$ is also the maximum likelihood estimator (MLE) for
 this problem, i.e.~the value of \eff\ which maximizes the likelihood
 function (\ref{eq-binomial}).  The MLE $f$ of \eff\ is an unbiased
 estimator with a number of attractive \emph{asymptotic} properties
 \cite{vandenbos2007,james06}, which justify its widespread use:
 \begin{itemize}
 \item Consistency: the MLE converges in probability to the true
       value;

 \item Asymptotic normality: the distribution of the MLE tends to
       the Gaussian distribution centred on the true value for very
       large sample sizes $n$;

 \item Efficiency: the MLE achieves the Cram\'er-Rao lower bound (no
       asymptotically unbiased estimator has lower asymptotic mean
       squared error than the MLE).
 \end{itemize}
 For the MLE $\hat\eff$ of a Bernoulli's process, the Cram\'er-Rao
 lower bound is
\[
   V(\hat\eff) 
     = \frac{1}{I(\eff)}
     = \frac{\eff(1-\eff)}{n} = \frac{V(k)}{n^2} = V(f) 
\]
 where 
\(
  I(\eff) = \sum_k \Bi(k|\eff,n)
            [ \partial\log\Bi(k|\eff,n) / \partial\eff]^2 
          = n [\eff(1-\eff)]^{-1}
\)
 is the Fisher information, and the last expression follows from the
 property $V(ax)=a^2 \,V(x)$, when $a$ is a known constant. 
 By replacing $\eff$ with the observed success frequency $f$ one
 obtains the widely (ab)used approximation
\begin{equation}\label{eq-approx-variance}
  V(\hat\eff) \approx f(1-f)/n = k(n-k)/n^3 \; .
\end{equation}

 The asymptotic properties are approximately valid also for moderately
 large values of \emph{both} $k$ and $n$, and this is the reason why
 the MLE $f$ and its approximate asymptotic uncertainty $\sigma_f =
 \sqrt{f(1-f)/n}$ are used so often.  However, they do not hold any
 more for small $n$ and when $k=0$ or $n$ (even if $n$ is large!).
 For example, the asymptotic variance is zero when $k=n$ (which makes
 sense for $n\to\infty$ because it means that the true value is
 $\eff=1$), but this is clearly a bad estimate of the uncertainty for
 any finite value of $n$: it would assign the same (zero) uncertainty
 to $k=n=1$ and $k=n=100$, even though one naively expects the latter
 result to be 10 times more precise than the former.  Finally, the
 confidence intervals $[f-\sigma_f, f+\sigma_f]$ have not always the
 correct coverage and, most important, may exceed the allowed
 boundaries of $\eff\in[0,1]$.

 To overcome these difficulties, several frequentist recipes have been
 proposed \cite{brown2001,cousins2010}, and the most common ones are
 available in ROOT \cite{root-cpc} as different options of the class
 TEfficiency.  Due to the discrete nature of the problem, obtaining
 the desired coverage for all possible values of $(n,\eff)$ is
 impossible.  The well known Clopper-Pearson confidence intervals
 never undercover hence they are to be preferred in conservative
 analyses, but are considered too wide by many experts, who proposed
 alternative recipes which provide the correct coverage on the
 average, allowing for some degree of under-coverage.  Usually, such
 approximations (illustrated in details in
 \cite{brown2001,cousins2010}) look quite similar in practical
 applications, and are often indistinguishable from the Bayesian
 reference credible intervals explained below.

 In HEP, people usually prefer to be conservative: wider confidence
 intervals which are known to never undercover are preferred in most
 situations.  However, while this can be critical in the Poisson case
 which is relevant for the search of new phenomena
 \cite{cousins0702156,casadei2012}, it may be argued that it is not as
 essential in the case of efficiency estimation.  When some
 approximate coverage is an accetable solution, as it looks reasonable
 for the specific case of interest here, then the choice is amongst a
 rather wide spectrum of alternative approximations.  As explained in
 details in ref.~\cite{cousins2010}, if Clopper-Pearson $q$-confidence
 intervals (which never undercover) are considered unacceptably too
 wide and one is ready to accept methods leading to intervals which
 may undercover, one may choose the Agresti-Coull approximation, which
 seldom undercovers, or approaches which give the desired average
 coverage like the Wilson formula or the Bayesian posterior reference
 $q$-credible intervals discussed below.  The last two recipes (Wilson
 and Bayesian intervals) give results which are numerically very
 similar: when they are considered acceptable approximations, our
 suggestion is to choose the Bayesian credible intervals because they
 can be interpreted intuitively in terms of the probability to contain
 the true unknown value.

\section{The Bayesian approach}\label{sec-Bayes-approach}

 In the Bayesian approach, the full solution is represented by the
 \emph{posterior} probability density\footnote{We use the lowercase
   $p$ for the probability density function and the uppercase $P$ for
   the probability distribution: $P(x) = \int^{x} p(t)\di t$.}
 $p(\eff|k,n)$ of the parameter of interest \eff\ (the selection
 efficiency), interpreted as our degree of belief about the possible
 values which \eff\ can assume.  The posterior density is obtained by
 means of the Bayes' theorem:
\begin{equation}
  p(\eff|k,n) = N \, \Bi(k|\eff,n) \, \pi(\eff)
\end{equation}
 in which the likelihood function is the binomial distribution
 $\Bi(k|\eff,n)$ from equation (\ref{eq-binomial}), viewed as a
 function of \eff, and the normalization constant is
\begin{equation}
  N = \frac{(n+1) \, B(k+1,n-k+1) \, B(a,b)}
           {B(a+k, b+n-k)}  \; .
\end{equation}
 Foundations require that the \emph{prior} density
 $\pi(\eff)$ encodes all information available about \eff\ before
 performing the experiment.  This kind of prior is often called
 ``subjective'' because it reflects the experimenter's degree of
 belief about different values based on the information available
 before performing the experiment.  Because such adjective is
 emotionally charged, here we prefer to call this kind of prior an
 ``informative'' prior instead, in contrast to the
 ``least-informative'' priors discussed below (which are elsewhere
 defined ``objective'').

 Because an informative prior is interpreted in terms of degree of
 belief, it must be a proper (i.e.~integrable) prior.  Whenever the
 experimenter does not feel very comfortable with a particular choice
 of prior, a sensitivity analysis has to be performed in which the
 solutions found by adopting different ``reasonable'' priors are
 compared.  In addition, one has the option of comparing the solutions
 corresponding to informative and least-informative priors.  In the
 large-sample limit, the final result will not depend significantly on
 the choice of the prior.  Hence, comparing the results obtained with
 different priors is also a way of assessing to what degree this limit
 is approached.

 If the prior is a function $\pi(\eff) = \Be(\eff | a,b)$ belonging to
 the Beta family (appendix~\ref{sec-Beta}), which contains the
 conjugate priors for the binomial model, the posterior also belongs
 to the same family:
\begin{equation}\label{eq-posterior}
  p(\eff|k,n) = \Bi(k| \eff, n) \, \Be(\eff | a,b)
              = \Be(\eff | k+a, n-k+b) 
\end{equation}
 This simplifies the math considerably.  Since (a) any regular
 function defined on $[0,1]$ can be obtained as a linear combination
 of Beta densities and (b) the linearity of the Bayes' theorem assures
 that in this case the posterior will be a linear combination
%
%
 of the corresponding posterior Beta densities
 \cite{BayesianTheory1994}, from the mathematical point of view one
 can limit herself to the Beta priors, as it will be done in the
 following.  This is of course by no means an excuse for not
 performing a sensitivity analysis: different priors can be treated
 either numerically or by expressing (or approximating) them as linear
 combinations of Beta densities.

\subsection{Informative priors}

 As stated above, the prior should represent all information available
 before performing the experiment.  A common case is when we only have
 limited prior information, for example some estimate of its
 expectation $E$ and variance $V$.  In this case, we can define an
 approximate prior by finding the Beta density with the same mean and
 variance with the method of moments: the Beta parameters $a,b$ are
 found by solving the two equations
\begin{eqnarray}
  a &=& E \left[ \frac{E (1-E)}{V} -1 \right]
  \label{eq-a-mom}
 \\
  b &=& (1-E) \left[ \frac{E (1-E)}{V} -1 \right]
  \label{eq-b-mom}
\end{eqnarray}
 Subtleties arise when the term in square brackets becomes negative,
 i.e.~when $E$ is very near to zero or one.  In this case, one can use
 an approximate prior whose parameters $a,b$ can be found from the
 formulae for the mode and variance (appendix~\ref{sec-Beta}).  The
 resulting Beta density is either monotonically decreasing with a
 maximum at zero or monotonically increasing with a maximum at one.  A
 unique Beta density is defined by $a,b$ found this way, hence one
 should also consider different choices of the prior (at least a
 least-informative prior) to check the sensitivity of the result.

 Another example is when the prior knowledge is the result of a
 different experiment.  In this case, the posterior density of the
 latter is used as the prior in equation (\ref{eq-posterior}) for the
 current experiment.\footnote{The other experiment does not need to be
   actually performed before our measurement.  ``Prior'' refers to our
   state of knowledge and does not imply any time ordering.}
 Then Bayes' theorem gives immediately the combined result of the
 two measurements. 

 A further example is the efficiency measurement performed with two
 runs carried on with the same accelerator and detector
 configurations.  Let $N=n_1+n_2$ be the total number of events
 collected in the two runs and $K=k_1+k_2$ be the total number of
 surviving events, where the outcome of run $j$ can be summarized in
 the sufficient statistic $(n_j,k_j)$.  With the prior $\Be(\eff|a,b)$
 used on the first run and the resulting posterior used as prior for
 the second run, one obtains:
\begin{equation*}
\begin{split}
  p(\eff|K,N) &= \Bi(k_2| \eff, n_2) \, \Be(\eff | k_1+a, n_1-k_1+b)
\\
      &= \Be(\eff | k_2+k_1+a, n_2-k_2 + n_1-k_1+b)
\\
      &= \Be(\eff | K+a, N-K+b)
\end{split}
\end{equation*}
 exactly the same result which corresponds to a single longer run
 defined as the union of both runs, as expected.

\subsection{Objective Bayesian results}\label{sec-ref-post}

 In HEP publications, it is often required to report results which
 only depend on the assumed model and on the observed data.  In this
 case, the choice of a ``least-informative'' prior which aims at
 encoding the minimal amount of information about the parameter is
 recommended.  The study of priors which guarantee ``objective''
 results in the sense precised above is the subject of the Bayesian
 \emph{reference analysis} \cite{bernardo05a}.  Such priors are called
 \emph{reference priors} and are formally defined such that they
 maximize the amount of missing prior information.  The definition
 makes only use of the asymptotic properties of the probability model
 \cite{berger09}.  The resulting \emph{reference posteriors} have the
 best coverage properties \cite{bernardo2007} in the sense that the
 posterior $q$-credible intervals obtained with any other prior
 achieve the coverage $q$ more slowly, for increasing sample sizes
 $n$.  In other words, reference priors are the ``probability
 matching'' priors whose $q$-credible intervals achieve the coverage
 $q$ most quickly for increasing sample sizes.

\begin{figure}[t!]
\begin{minipage}[b]{0.5\textwidth}
  \centering
  \includegraphics[width=\textwidth]{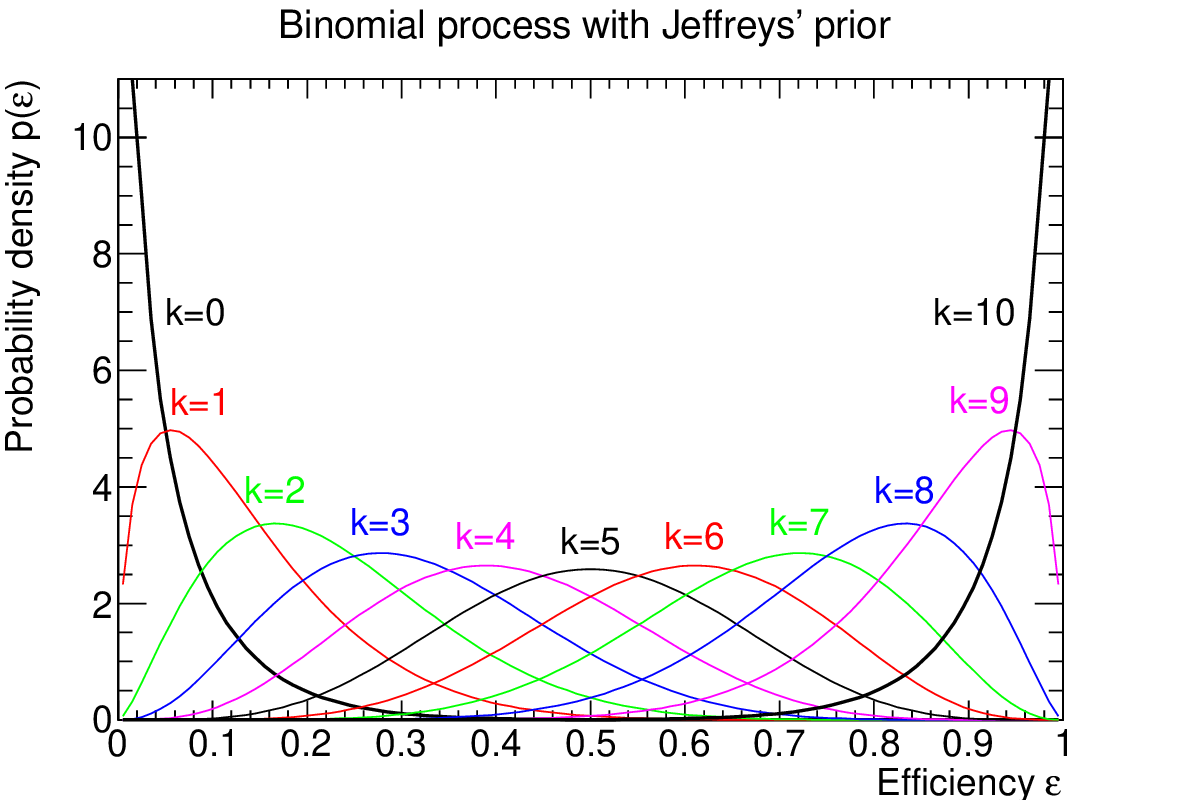}
\end{minipage}%
\begin{minipage}[b]{0.5\textwidth}
  \centering
  \includegraphics[width=\textwidth]{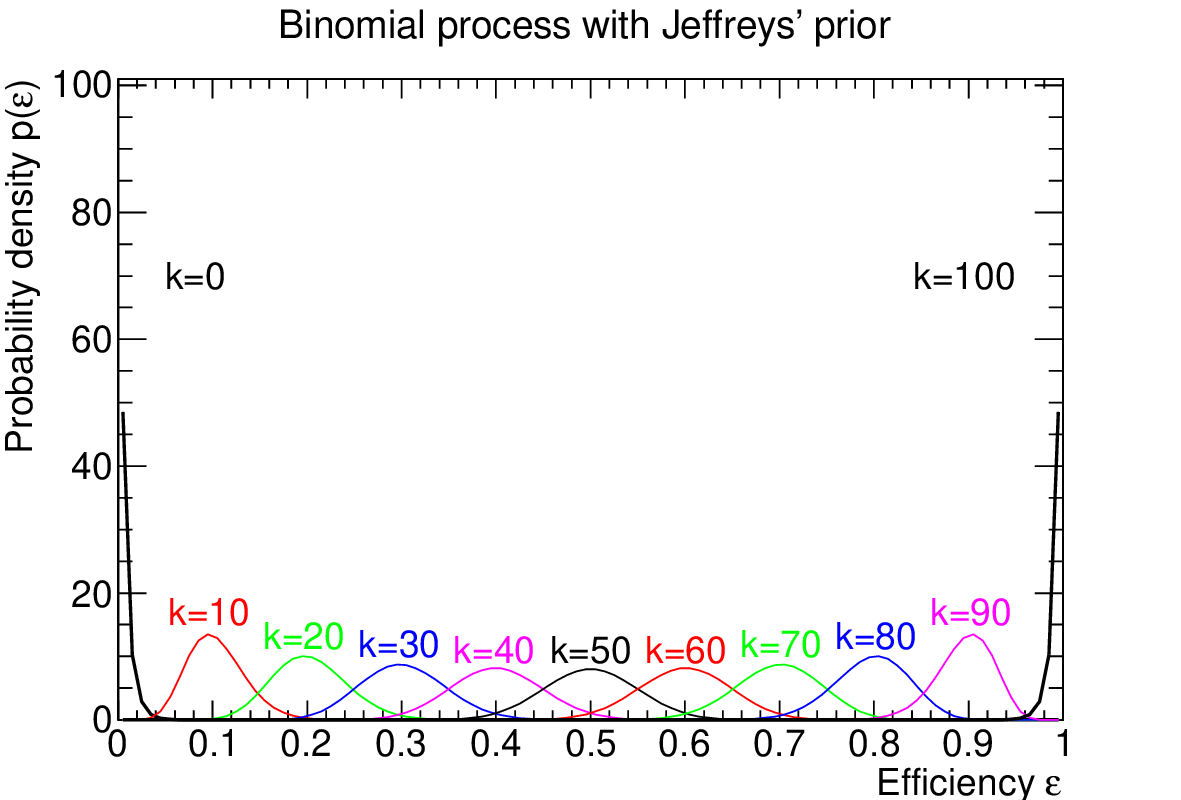}
\end{minipage}%
  \caption{Reference posterior probability density function
    $\Be(\eff;k+0.5,n-k+0.5)$ for $n=10$ (left) and $n=100$
    (right).}
  \label{fig-pdf-Jeffreys-prior}
\end{figure}

 For the binomial model\footnote{For the negative binomial model the
   reference prior is $\text{Be}(\eff|0,\frac{1}{2}) \propto
   \eff^{-1} (1-\eff)^{-1/2}$.}, the reference prior coincides with the
 Jeffreys' prior \cite{jeffreys46} and is $\pi(\eff) =
 \text{Be}(\eff|\frac{1}{2},\frac{1}{2}) \propto \eff^{-1/2}
 (1-\eff)^{-1/2}$, which has two maxima at the extrema and a unique
 minimum at $\eff=1/2$.  This means that the reference posterior
 density for the unknown parameter is
\begin{equation}
  p(\eff|r,n) = \text{Be}(\eff|r+\tfrac{1}{2}, n-r+\tfrac{1}{2})
    \propto \eff^{r-1/2} (1-\eff)^{n-r-1/2}
\end{equation}
 which has mean $E(\eff)=(r+\frac{1}{2})/(n+1)$, a biased estimator of
 the true efficiency, and variance $V(\eff) =
 (r+\tfrac{1}{2})(n-r+\tfrac{1}{2})/[(n+1)^2 (n+2)]$.  The variance
 for $r=0$ or $r=n$ decreases as $n^{-2}$ for large sample sizes $n$,
 as one naively expects, hence does not pose the problems arising from
 the use of the asymptotic expression derived from the MLE variance,
 equation~(\ref{eq-approx-variance}), although the two expressions
 converge for large sample sizes.  Figure~\ref{fig-pdf-Jeffreys-prior}
 shows the reference posterior densities for a small sample ($n=10$,
 left plot), which shows a clear asymmetry in most cases, and a
 moderately large sample ($n=100$, right plot), for which the (symmetric)
 asymptotic expression provides a good approximation whenever the
 observed frequency is not too near zero or one.

 Quite often, people have used the uniform prior $\Be(\eff|1,1)$ in
 place of Jeffreys' prior stating that the uniform prior is ``non
 informative''.  However this is not true in the binomial case,
 although one can find a 1:1 transformation $\psi=\psi(\eff)$ such
 that the reference prior of the transformed variable $\psi$ is
 uniform (the \emph{reference parametrization}) \cite{bernardo05a}:
 the transformation is $\psi(\eff) = \frac{2}{\pi} \arcsin\sqrt{\eff}$
 whose inverse is $\eff(\psi)=\sin^2(\psi\pi/2)$.  In terms of the
 original parameter $\eff$, the uniform prior is to be considered an
 informative prior.  Incidentally, one can notice that the posterior
 mode is equal to the observed success frequency in this case,
 i.e.~the posterior peak coincides with the MLE, because the whole
 posterior coincides with the likelihood function.

\begin{figure}[t!]
\begin{minipage}[b]{0.5\textwidth}
  \centering
  \includegraphics[width=\textwidth]{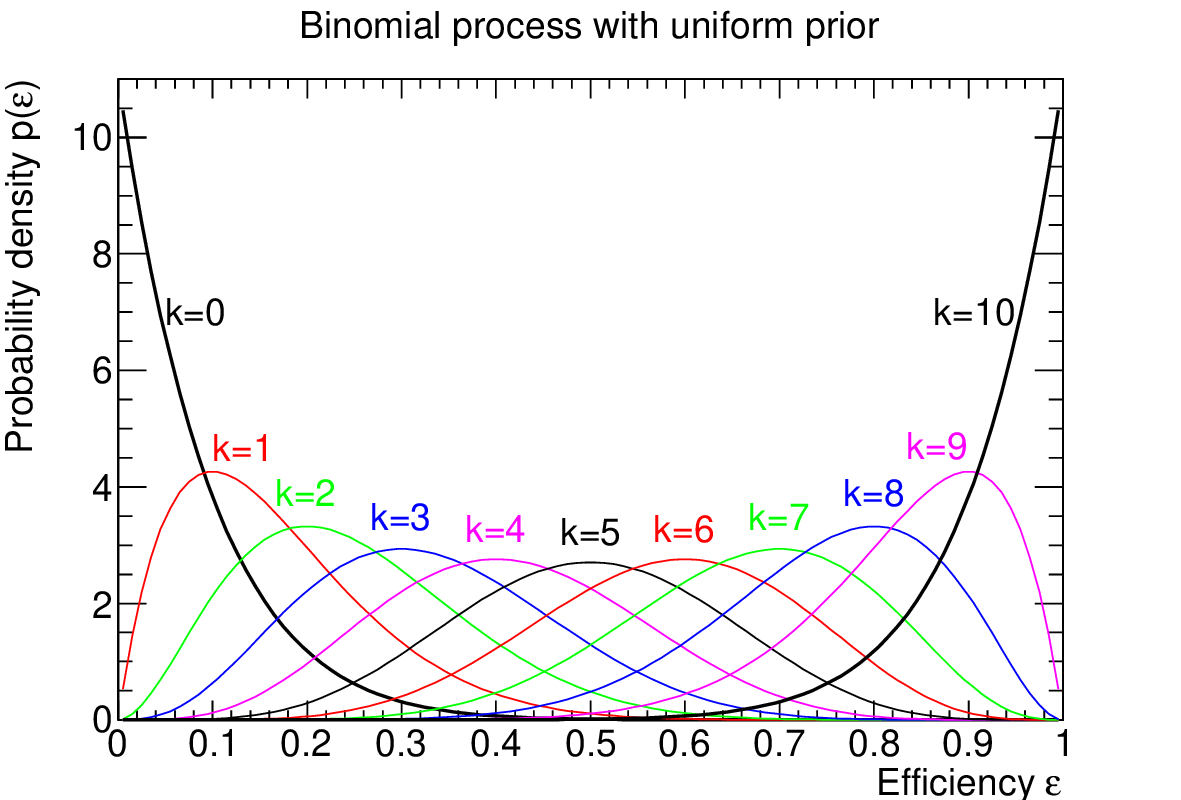}
\end{minipage}%
\begin{minipage}[b]{0.5\textwidth}
  \centering
  \includegraphics[width=\textwidth]{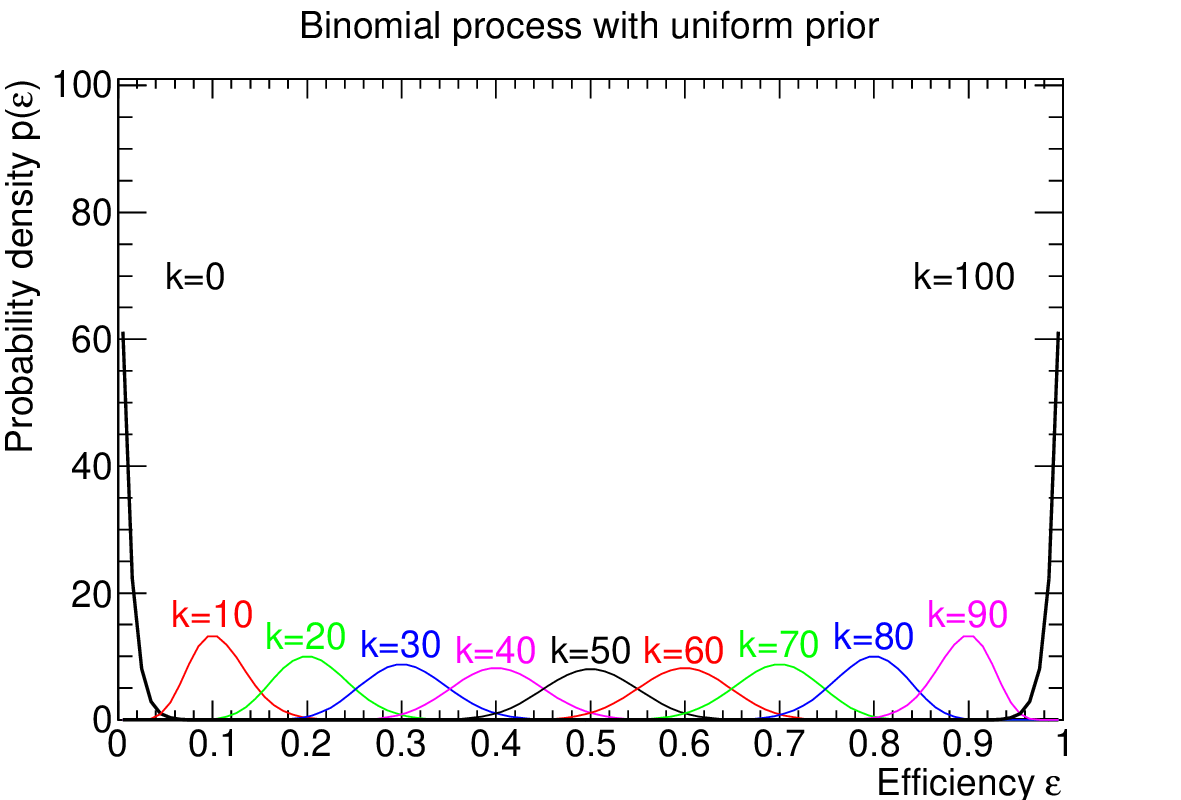}
\end{minipage}%
  \caption{Posterior density $p(\eff|k,n) = \Be(\eff;k+1,n-k+1)$
    obtained with the uniform prior, for $n=10$ (left) and $n=100$
    (right).}
  \label{fig-pdf-unif-prior}
\end{figure}

 Our recommendation is to choose the reference prior when aiming at
 reporting results which only depend on the assumed model (the
 binomial process) and the observed data.  However, the parameters of
 the posterior Beta density differ only by half unit when using the
 reference or the uniform prior.  Hence in practice (unless the sample
 size is very small) there is not a big difference between the two
 posteriors.  In particular, also the variance obtained with a uniform
 prior does not suffer from the problems of the asymptotic expression
 and gives similar results to the reference posterior variance already
 for $n=10$, as it appears from the comparison between
 figure~\ref{fig-pdf-Jeffreys-prior}, showing reference posteriors,
 and figure~\ref{fig-pdf-unif-prior}, showing the result obtained with
 a uniform prior for the same sample sizes.

 The reference posterior mean and variance can be used in computations
 involving the efficiency, for example when estimating a cross
 section.  The usual algebra of variances hold in this case, hence one
 can compute the best estimate of the cross section and its variance
 in the usual way.  However, the posterior mean is not the only
 possibility for the ``best'' estimate.  In addition, the posterior
 credible intervals are usually not symmetric about the mean.  When
 these aspects are important, one should take into account what is
 treated in the following section.

\subsection{Bayesian inference}\label{sec-intrinsic-estimators}

 In the Bayesian framework, the statistical inference is treated as a
 decision problem \cite{Bernardo2003} in which one chooses the
 estimate which minimizes the posterior expected loss, for a suitably
 chosen loss function.  Clearly, a very desirable property for an
 estimator $\eff^*$ is the invariance under reparametrization, in the
 sense that the best estimator of a 1:1 function $\phi=\phi(\eff)$ is
 $\phi^*=\phi(\eff^*)$.  This is not achieved by the widespread
 quadratic loss, for which the best estimate is the posterior mean,
 because the quadratic loss is not invariant under reparametrization.

 An example of invariant loss function is the $\mathcal{L}_1$ norm,
 that is the integral of the absolute value of the difference between
 two distributions, computed at the same point over the whole support.
 When applied to the reference posterior for the binomial case, the
 $\mathcal{L}_1$ norm gives the invariant expected loss
\begin{eqnarray}\label{eq-l1-norm}
 \ell_1\{\eff_0|k,n\} &=& \int_0^1
    \ell_1(\eff_0,\eff) \,
    \Be(\eff|k+\tfrac{1}{2},n-k+\tfrac{1}{2}) \di\eff
\\
 \ell_1(\eff_0,\eff) &=& \sum_{k=0}^n
    \left| \, \Bi(k|\eff_0,n) - \Bi(k|\eff,n) \, \right|
\end{eqnarray}
 independent from one-to-one transformations of $\eff$.  One can build
 a \emph{lowest posterior loss} (LPL) $q$-credible region by finding
 the interval $[\eff_{\text{low}}, \eff_{\text{up}}] \subset [0,1]$
 which minimizes (\ref{eq-l1-norm}) under the constraint \(
 \int_{\eff_{\text{low}}}^{\eff_{\text{up}}}
 \Be(\eff|k+\tfrac{1}{2},n-k+\tfrac{1}{2}) \di\eff = q \).

 The behaviour of many important limiting processes in probability
 theory and statistical inference is better described in terms of
 another measure of divergence, related to the information theory, the
 \emph{intrinsic discrepancy} $\delta\{p_1,p_2\}$, defined
 as the minimum among the two Kullback-Leibler \emph{directed
 divergences} between two probability models $p_1$ and $p_2$
 \cite{Bernardo2003}:
\begin{eqnarray}
 \delta\{p_1,p_2\} &=& \min\left\{ \kappa\{p_1,p_2\} ,
                                   \kappa\{p_2,p_1\} \right\}
\\
 \kappa\{p_i,p_j\} &=& \int_{\mathbb{X}} p_i(\vec{x})
       \log\frac{p_i(\vec{x})}{p_j(\vec{x})} \di \vec{x}
\end{eqnarray}
 The intrinsic discrepancy is symmetric, non-negative, defined for
 strictly nested supports, invariant under one-to-one transformations,
 and additive for independent observations.  It may be viewed as the
 minimum expected log-likelihood ratio in favour of the model which
 generates the data (the ``true'' model, which is assumed to be
 described either by $p_1$ or $p_2$) and can be used to defined the
 \emph{intrinsic discrepancy loss}
\begin{equation}
 \delta_{\vec{x}} \{\vec{\theta}_0, \vec{\theta}\} = \delta\{p(\vec{x}|\vec{\theta}_0), p(\vec{x}|\vec{\theta})\}
\end{equation}
 where $\vec{\theta}$ is the parameter in which we are interested. 

 For the binomial model considered here $\vec{\theta}=\eff$ and the
 intrinsic discrepancy loss is
\begin{eqnarray}
 \delta_k\{\eff_0,\eff|n\} &=&
       n \, \delta\{\eff_0,\eff\}
\\
 \delta\{\eff_0,\eff\} &=& 
       \min\left\{ \kappa\{\eff_0|\eff\} ,
                   \kappa\{\eff|\eff_0\} \right\}
\\
 \kappa\{\eff_i|\eff_j\} &=&
       \eff_j \log\frac{\eff_j}{\eff_i}
      + (1-\eff_j) \log\frac{1-\eff_j}{1-\eff_i}
\end{eqnarray}
 where $\delta\{\eff_0,\eff\}$ is the intrinsic discrepancy between
 Bernoulli's random variables with parameters $\eff_0$ and $\eff$.

 The reference posterior expectation of the intrinsic discrepancy loss
 is the (\emph{reference posterior}) \emph{intrinsic loss}.  The value
 $\vec{\theta}^*$ which minimizes the intrinsic loss is a Bayesian estimator
 which is called the \emph{intrinsic estimator} of $\vec{\theta}$, and the
 reference posterior $q$-credible region which minimize the intrinsic
 loss is the \emph{intrinsic $q$-credible region} of $\vec{\theta}$.  Such
 credible regions are invariant under reparametrization and always
 contain $\vec{\theta}^*$, which is also invariant.  In addition, the
 intrinsic $q$-credible regions are always approximate confidence
 regions with coverage $q$ and in some case they have the exact
 coverage, as it happens for location-scale models \cite{bernardo2007}.

 The intrinsic loss for the binomial model is
\begin{equation}\label{eq-ref-post-exp-loss}
  d(\eff_0|k,n) = n \int_0^1 \delta_x\{\eff_0,\eff\}
    \, \Be(\eff|k+\tfrac{1}{2},n-k+\tfrac{1}{2}) \di\eff
\end{equation}
 and the intrinsic estimator $\eff^*$ is the value which minimizes
 equation (\ref{eq-ref-post-exp-loss}).  The intrinsic $q$-credible
 interval is the interval $[\eff_a(k,n), \eff_b(k,n)]
 \subset [0,1]$ which minimizes the loss~(\ref{eq-ref-post-exp-loss})
 under the constraint
\(
  \int_{\eff_{\text{low}}}^{\eff_{\text{up}}}
    \Be(\eff|k+\tfrac{1}{2},n-k+\tfrac{1}{2})
      \di\eff = q
\)
 (a simple numerical algorithm in suggested in
 appendix~\ref{app-intrinsic-intervals}).  For the binomial model,
 because of its discrete nature, the exact coverage is not achieved for
 finite sample sizes, similarly to the confidence intervals obtained
 with the frequentist approach.  One of the advantages of summarizing
 the result by providing $\eff^*$ and the intrinsic interval $[\eff_a,
 \eff_b]$ is that they are invariant under 1:1 reparametrization, in the
 sense that the intrinsic estimator of $\phi(\eff)$ is $\phi(\eff^*)$
 and its intrinsic $q$-credible interval is the $\phi$-image
 $\phi([\eff_a, \eff_b])$ of $[\eff_a, \eff_b]$.  Hence, they are the
 recommended summaries of the full Bayesian solution (which is the
 reference posterior) when aiming at reporting an ``objective''
 result.

 A numerical treatment is needed to find the intrinsic estimator and
 the intrinsic credible intervals for the binomial model.  However,
 one can obtain approximations good already with moderate sample sizes
 (say $n$ larger than few tens) by considering the approximate
 location parameter $\psi(\eff) = \frac{2}{\pi} \arcsin\sqrt{\eff}$
 \cite{bernardo05a}, such that the intrinsic estimator is
\begin{equation}\label{eq-approx-intrinsic-estimator}
  \eff^*(k,n) \approx \sin^2 \left(\frac{\pi}{2} E[\psi|k,n]\right)
              \approx \frac{k+\frac{1}{3}}{n+\frac{2}{3}}
\end{equation}
 (a very good approximation) and the reference posterior intrinsic
 loss is
\begin{equation}\label{eq-approx-intrinsic-loss}
  d(\eff_0|k,n) \approx \frac{1}{2} + 2 n \left[
   \arcsin\sqrt{\eff_0} -
   \arcsin\sqrt{(k+\alpha_n)/(n+2\alpha_n)}
  \right]^2
\end{equation}
 where $\alpha_n=(n+4)/(4n+10)$ converges to $1/4$ for large $n$.

 Simple (although less accurate) approximate expressions for the
 intrinsic credible intervals can be built upon the parametrization
 $\eff(\phi) = \sin^2 (\phi/2)$ \cite{bernardo05b}.  Using a shorter
 notation for the reference posterior mean
 $\mu=E(\eff)=(k+\tfrac{1}{2})/(n+1)$ and variance
 $\sigma^2=V(\eff)=\mu(1-\mu)/(n+2)$ of the parameter of interest, the
 variance of the reference parametrization is
\begin{equation}
  \sigma_\phi^2 \approx \sigma^2 [\phi'(\mu)]^2
   = \frac{1}{n+2}
\end{equation}
 while its mean is
\begin{equation}\label{eq-ref-par-intrinsic-estimator}
\begin{split}
  \mu_\phi &\approx \phi(\mu) + \frac{1}{2} \sigma^2 \phi''(\mu)
\\
   &= 2\arcsin\sqrt{\frac{k+\tfrac{1}{2}}{n+1}} + \frac{(2k-n)}{4(n+2)}
       \left[(k+\tfrac{1}{2}) (n-k+\tfrac{1}{2}) \right]^{-1/2} 
\end{split}
\end{equation}
 where $\phi'$ and $\phi''$ denote the first and second derivative
 with respect to $\eff$.  Finally, the asymptotic intrinsic
 $q$-credible interval in the reference parametrization is
 $[\phi_{-},\phi_{+}]$ where
\begin{equation}\label{eq-ref-par-intrinsic-interval}
  \phi_{\pm} \approx \mu_\phi \pm z_q \sigma_\phi
   = \mu_\phi \pm \frac{z_q}{\sqrt{n+2}}
\end{equation}
 where $z_q$ is the $(q+1)/2$ quantile of the normal distribution.
 The intrinsic $q$-credible interval for the efficiency is obtained by
 transforming back to $\eff_{\pm} = \eff(\phi_{\pm}) =
 \sin^2 (\phi_{\pm}/2)$.

\begin{figure}[t!]
  \centering
  \includegraphics[width=\textwidth]{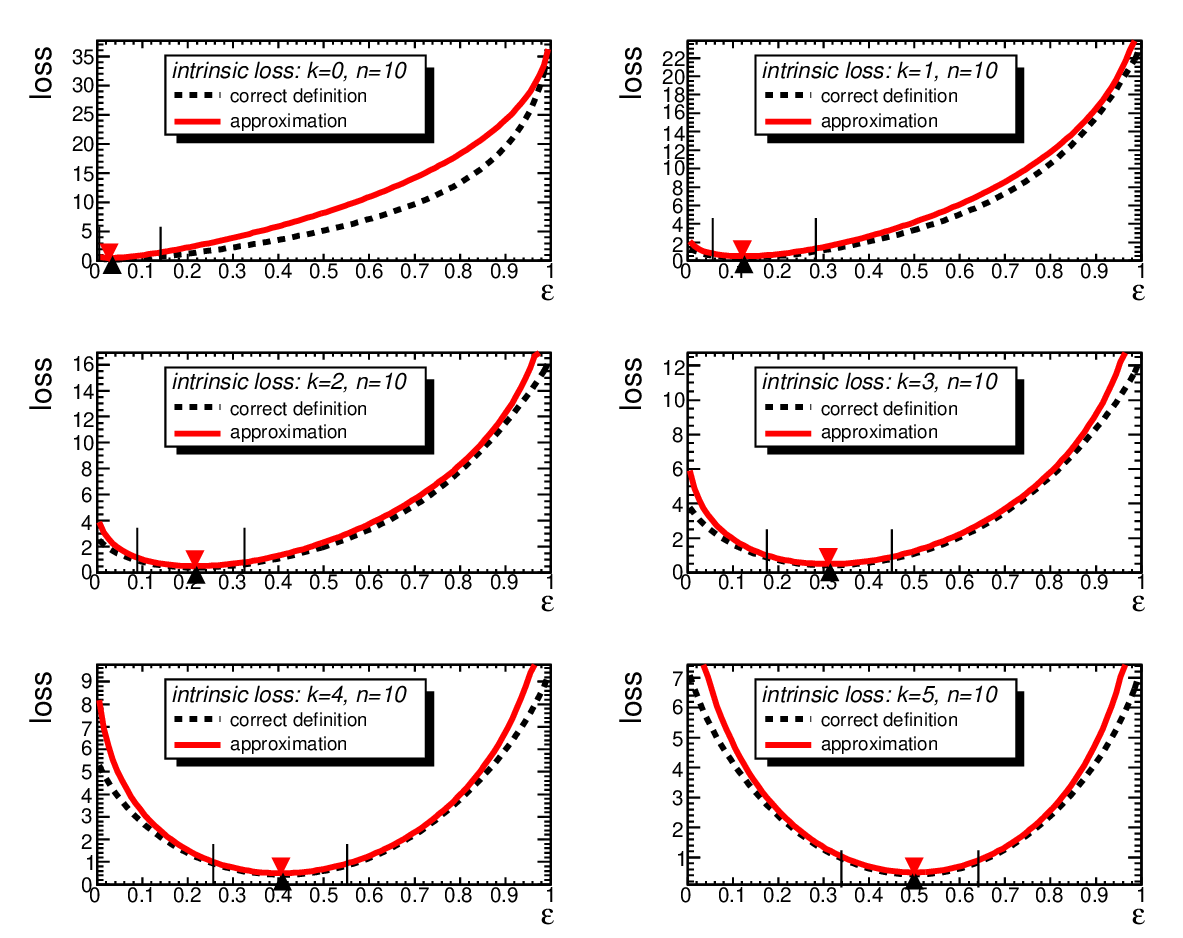}
  \caption{Reference posterior intrinsic loss from
    the reference posterior expected loss
    (black dashed line) and its approximation
    (continuous red line) for a small sample size $n=10$.  The
    triangles show the position of the minima of such functions and
    the vertical lines represent the limits of the 68.3\% intrinsic
    credible interval (which becomes automatically one-sided in the
    case $k=0$).  }
  \label{fig-intrinsic-loss}
\end{figure}

\begin{table}[t!]
\footnotesize
\centering
\begin{tabular}{c|ccc|cc|cc}
 &
 \multicolumn{3}{c|}{\textbf{Exact}} &
 \multicolumn{2}{c|}{\textbf{Approx 1}} &
 \multicolumn{2}{c}{\textbf{Approx 2}} \\
 $k/n$ & $d_{\text{min}}$ & $\eff^*$ & 68,3\% interv. &
 $\eff(\mu_{\phi})$ & 68,3\% interv. &
 $\eff^{*}_{\text{app}}$ & 68,3\% interv. \\
 \hline
 0/10 & 0.24 & 0.033 & [0.000, 0.060] &  0.028 & [0.002, 0.082] &  0.031 & [0.000, 0.060] \\
 1/10 & 0.36 & 0.124 & [0.024, 0.193] &  0.122 & [0.054, 0.211] &  0.125 & [0.024, 0.193] \\
 2/10 & 0.40 & 0.218 & [0.090, 0.326] &  0.216 & [0.125, 0.323] &  0.219 & [0.090, 0.326] \\
 3/10 & 0.42 & 0.314 & [0.171, 0.446] &  0.311 & [0.205, 0.427] &  0.313 & [0.169, 0.443] \\
 4/10 & 0.43 & 0.408 & [0.257, 0.552] &  0.405 & [0.290, 0.526] &  0.406 & [0.257, 0.552] \\
 5/10 & 0.43 & 0.500 & [0.350, 0.651] &  0.500 & [0.380, 0.620] &  0.500 & [0.350, 0.651] \\
\end{tabular}
\caption{For a small sample size $n=10$, the minimum value of the
  reference posterior intrinsic loss, the intrinsic estimator and the
  intrinsic 68.3\% credible interval from the exact definition
  eq.~(3.13) 
  are reported. 
  Next, the approximated quantities derived from the reference
  parametrization $\eff(\mu_{\phi})$ from
  eq.~(3.17) 
  and $\eff(\phi_{\pm})$ from
  eq.~(3.18) 
  are shown.
  Finally, the approximate  intrinsic estimator
  $\eff^{*}_{\text{app}} = (k+1/3)/(n+2/3)$ from
  eq.~(3.14) 
  and the 68.3\% credible interval
  from the approximate intrinsic loss defined by 
  eq.~(3.15) 
  are reported, both representing a very good approximation.
}
\label{tab-intrinsic-loss}
\end{table}

 Figure~\ref{fig-intrinsic-loss} shows the comparison between the
 correct [eq.~(\ref{eq-ref-post-exp-loss})] and approximate
 [eq.~(\ref{eq-approx-intrinsic-loss})] intrinsic loss functions and
 table~\ref{tab-intrinsic-loss} reports the results obtained with them
 and with the reference parametrization.  A small sample size $n=10$
 is chosen such that these approximations give different results when
 reporting three decimal places
 and the values corresponding to $k=6,\ldots,10$ are omitted because
 they are symmetrical with respect to those obtained for
 $k=0,\ldots,5$.  The overall agreement is fairly good even with such
 small sample.

 The value $\eff^{*}_{\text{app}} = (k+1/3)/(n+2/3)$ which minimizes
 the approximate intrinsic loss defined by
 eq.~(\ref{eq-approx-intrinsic-loss}) provides a very good
 approximation to the exact intrinsic estimator $\eff^*$ and the
 intervals (numerically) computed from the approximate intrinsic loss
 are practically the same as the exact intrinsic intervals, despite
 from the very small sample size $n=10$.  The approximate quantities
 computed from the mean $\mu_{\phi}$ of the reference parametrization
 [eq.~(\ref{eq-ref-par-intrinsic-estimator}] and from its approximate
 credible intervals from eq.~(\ref{eq-ref-par-intrinsic-interval}) are
 less good for such a small sample size, but become better with larger
 $n$.  If their accuracy is considered acceptable, the approximate
 credible intervals from eq.~(\ref{eq-ref-par-intrinsic-interval}) may
 be used together with $\eff^{*}_{\text{app}}$ to provide a summary of
 the posterior which does not require a numerical algorithm to be
 developed to minimize the loss.

 As mentioned above, the binomial model is discrete, hence the
 coverage is never exact.  The intrinsic $q$-credible intervals are
 also approximate confidence intervals with coverage $q$, which may
 undercover or overcover depending on the actual values of $(n,\eff)$,
 similarly to the approximate frequentist confidence intervals
 reviewed in \cite{cousins2010}.  This justifies the use of the
 approximate intrinsic intervals already for moderate sample sizes in
 all the cases in which an approximate coverage is accepted (that is
 when the Bayesian approach is chosen or when the frequentist
 approximate confidence intervals are considered acceptable
 approximations).


\section{Non trivial use cases}\label{sec-patologic}

 So far, we assumed that all entries of the initial histogram had unit
 weight and that the events had been selected by an independent
 process, whose efficiency is completely uncorrelated with respect to
 the efficiency of the process under study (which means that it is
 expected to reduce the total number of events and the number of
 selected events in such a way that the long run success frequency
 remains unchanged).  This is required to obtain a binomial process,
 but may not be true in all cases, as it happens sometimes in HEP
 problems.

 For example, if the initial sample was obtained by scaling the
 simulated data sample to normalize it to some different value of the
 cross section, it \emph{should not be used} to make efficiency
 studies!  Rather, the efficiency should be estimated by using the
 \emph{original} sample (with unit weights), in order to have a
 binomial process.  Once the efficiency is known, it may be applied to
 the scaled distribution to estimate the total number of surviving
 events.

 The following common examples are addressed in detail below.
\begin{enumerate}
  \item Sometimes one is interested into the overall selection
    efficiency for a weighted mixture of different samples.  This is a
    common use case in HEP, because Monte Carlo (MC) generators are
    often used to produce independent samples for different processes
    which greatly differ in their cross section, and is addressed
    below in section~\ref{sec-weights}.  Another common example in HEP
    is the output of MC@NLO \cite{mcatnlo}, a mixture of events
    with weights $\pm1$ (section~\ref{sec-posneg-weights}).

  \item When measuring the trigger efficiency, the initial sample
    might be selected by a non independent process.  This happens for
    example when a random trigger provides not enough events to allow
    for studying the efficiency of the trigger of interest, and is
    addressed in section~\ref{sec-trigger}.
\end{enumerate}

 The most important assumption in the efficiency estimation methods
 considered here is that each individual event can be considered as a
 Bernoulli trial which contributes with either zero or one success to
 the count of surviving events.  This gives the binomial\footnote{Or
   negative-binomial, although we are not interested into it here.}
 likelihood, which is the starting point for both fequentist and
 Bayesian methods.
%
%
 The only case in which weighted events can be reconduced to the
 binomial scheme is when only integer weights are allowed, with the
 interpretation that an event with weight $m$ is just a short-hand
 \emph{notation} for a set of $m$ Bernoulli's trials \emph{all} either
 passing or failing the selection.  In this case, the likelihood for
 such events is just the product of $m$ identical likelihoods, that is
 the $m$-th power of the Bernoulli's model.  Moving from integer to
 real weights by assuming that the form of the likelihood is unchanged
 requires a mathematical proof which is unknown to the author.

 In section~8.5 of James' book \cite{james06} an approximate
 likelihood is described which uses real weights but requires quite
 some care because of its ``paradoxical results''.  One delicate point
 of his approach is that the variance of the estimator may be reduced
 by dropping events (in particular, those with large weights).
 The TEfficiency class offered by the ROOT framework implements it,
 allowing for arbitrary event weights, possibly dependent on some
 event property or observable quantity.  However, this is not
 considered further in this article, in which we restrict the problem
 to a smaller set of use cases (in which the solutions proposed below
 are practically equivalent to the result obtained with TEfficiency).

 Here we only consider situations in which we have a finite number $s$
 of possible event sets (or classifications), and the event weights
 only depend on $s$, which implies that they can only assume $s$
 possible values $w_1, \ldots, w_s$ with probabilities $p_i$
 satisfying $\sum_{i=1}^{s} p_i = 1$.  More explicitly, we do not
 consider weights which depend on some continuous observable quantity,
 nor the case of infinite countable weights.

 The general case of unknown classification (which holds for the
 experimental data) cannot be fully solved, even when the
 probabilities $p_i$ (associated to all possible $s$ channels) are all
 known.  In this case the sample consists of $N$ events and one has a
 multinomial distribution which relates the sizes $n_1+\ldots+n_s=N$
 of the $s$ categories to the corresponding probabilities
 $p_1,\ldots,p_s$, and a binomial term for each category which relates
 the number $k_i$ of events surviving the selection out of the initial
 $n_i$ events to the subsample efficiency $\eff_i$.  In other words,
 the likelihood is proportional to the product
\begin{equation}\label{eq-multinomial-binomial}
  P(k_1,\ldots,k_s; n_1,\ldots,n_s; \eff_1,\ldots,\eff_s) \propto
   \frac{N! \, p_{1}^{n_1} \cdots p_{s}^{n_s}}{n_{1}! \dots n_{s}!}
   \,
   \prod_{i=1}^{s} \binom{n_i}{k_i} \eff^{k_i} (1-\eff)^{n_i-k_i}
\end{equation}
 of a multinomial distribution which gives the probability of each
 partition $n_1+\ldots+n_s=N$ and $s$ binomial terms which express the
 probability of selecting all possible $s$-tuples $(0 \le k_1 \le
 n_1$, $\ldots$, $0 \le k_s \le n_s)$ such that $\sum_{i=1}^{s} k_i =
 K$.  Because the only observable numbers are the initial sample size
 $N$ and the number $K$ of events passing the selection, even after
 the sum over all possible $s$-tuples $(n_1,\ldots,n_s)$ and
 $(k_1,\ldots,k_s)$, we are left with a $s$-dimensional problem in
 which the efficiencies $\eff_1,\ldots,\eff_s$ cannot be determined,
 unless they satisfy some relation known a priori which allows to
 remove (or integrate over) $s-1$ degrees of freedom.

 A common practical problem is the combination of several MC samples
 which correspond to independent physical processes, with known
 cross-sections.  In this situation, the classification of each event
 and the probabilities $p_i$ are known: the file name provides the
 classification and both the weight $w_i$ and the probability $p_i
 = w_i / \sum_{i=1}^{s} w_i$ are proportional to the cross-section.
 Hence the general likelihood (\ref{eq-multinomial-binomial}) is
 reduced to the product of independent binomial factors with known
 initial ($n_i$) and final ($k_i$) numbers of events and unknown
 efficiencies.  This splits into a collection of individual problems:
 we can estimate the selection efficiency for each possible class as
 explained above, and compute the overall efficiency for the weighted
 mixture as illustrated in section~\ref{sec-weights} below.

 In the particular case of MC@NLO we have $s=2$ categories to which
 any event can be associated thanks to the weights $w_1 \equiv w_+ =
 +1$ and $w_2 \equiv w_- = -1$ assigned by the generator.  From the
 user perspective the corresponding probabilities $p_\pm$ are unknown,
 although they can be estimated by looking at the ratio between the
 number $n_\pm$ of events in each category and the total number $n_+ +
 n_-$ of generated events.  Anyway, we are not interested into the
 values of $p_\pm$ themselves hence we can integrate
 (\ref{eq-multinomial-binomial}) over them.  This way, the marginal
 model has a likelihood function which only depends on the
 single-sample efficiencies $\eff_{\pm}$ and this becomes a particular
 case of the problem mentioned above (more details on
 section~\ref{sec-posneg-weights} below).

\subsection{Overall efficiency for mixed samples}\label{sec-weights}

 A frequently encountered problem in HEP occurs when simulating
 concurrent processes with different cross sections.  Interesting
 observables like the transverse momentum of electrons or jets may
 have distributions which decrease so rapidly that producing fully
 simulated samples with enough events in the whole interesting range
 (especially in the tail) is impractical, unless one splits the
 generator-level input into different samples with roughly the same
 size but different ranges (for example starting from a \Pt\ value
 which doubles in each sample).  In order to obtain a smooth
 distribution, such samples need to be mixed with weights which are
 proportional to their relative cross-sections.

 Let us consider a finite number $s$ of independently generated
 samples with sizes $N_i$, $i=1,\ldots,s$ whose weights $w_i>0$ are
 perfectly known.  Before the selection, the effective sample size is
\[
    N = \sum_i w_i N_i
\]
 where $N_i$ is the (known) initial size of the $i$-th sample.

 Let $K_i$ be the known sample size after the selection and $\eff_i$
 be the unknown selection efficiency for the $i$-th sample.  Then, the
 effective sample size after the selection is
\[
    K = \sum_i w_i K_i = \sum_i w_i N_i \eff_i
\]
 and we are interested into the overall selection efficiency for the
 mixed sample.

 Because all samples are independent, one can estimate the efficiency
 $\eff_i$ and its variance $v_i$ for each sample separately, as
 explained in the previous sections.  Next, the variance of $K$ can be
 computed with the usual algebra of variances:
\begin{equation}
      V(K) = \sum_i w_i^2 \, V(K_i) = \sum_i (w_i N_i)^2 v_i  \; .
\end{equation}

 When considering the mixed sample, one can estimate the unknown true
 efficiency by taking the ratio $E$ between $K$ and $N$:
\begin{equation}\label{eq-effective-eff}
    E = \frac{K}{N} = \frac{\sum_i W_i \eff_i}{\sum_i W_i}
\end{equation}
 which is a linear combination of the single-sample efficiencies
 $\eff_i$ with weights $W_i = w_i N_i$.  Its variance is
\begin{equation}\label{eq-effective-var}
    V = \left(\sum_i W_i^2 v_i\right) / \left(\sum_i W_i\right)^2
    \;.
\end{equation}

\begin{figure}[t!]
  \centering
  \includegraphics[width=0.75\textwidth]{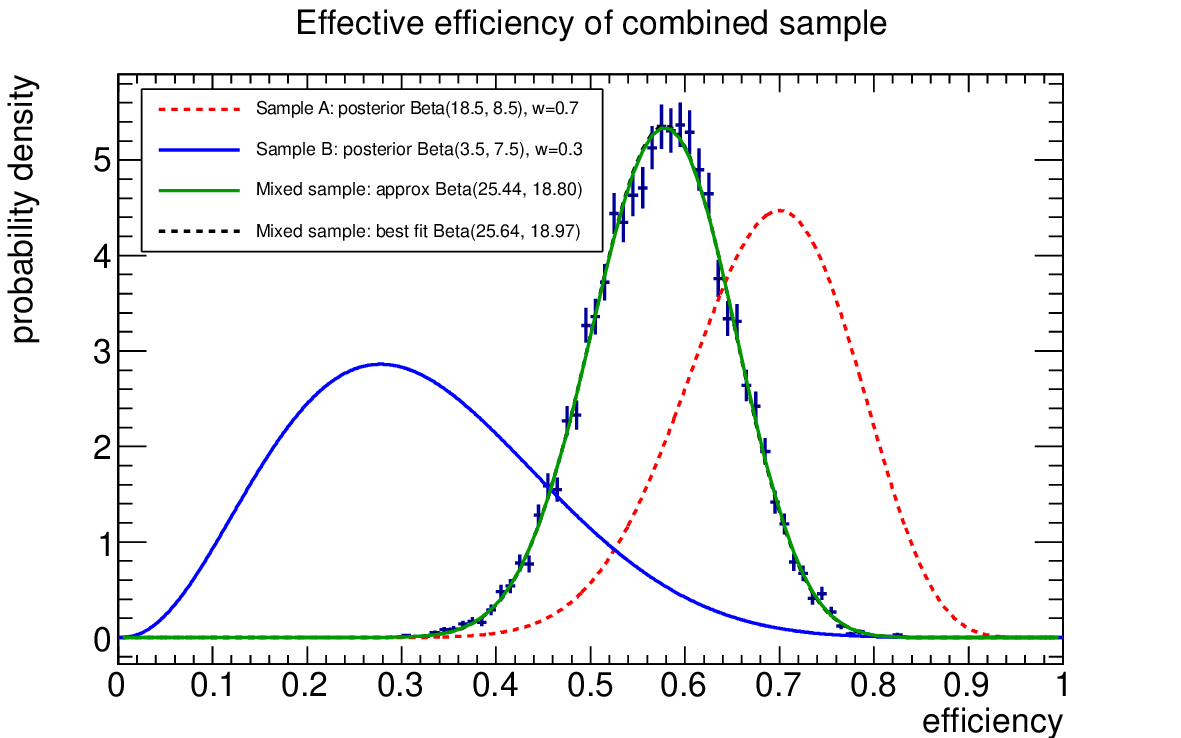}
  \caption[Overall efficiency.]{The overall efficiency for a
    selection carried on with a mixture of two samples with weights
    70\% and 30\% and single-sample (true) efficiencies of 70\% (18
    selected out of 26 events) and 30\% (3 selected out of 10 events)
    is estimated by performing 10000 repeated pseudo-experiments.  The
    differences between the best fit and the approximate Beta density
    is not noticeable from the plot.}
 \label{fig-betas}
\end{figure}

 Instead of reporting $E \pm \sqrt{V}$, which would cause troubles if
 $E$ is too near zero or one, we may find the approximate asymmetric
 uncertainty band around the estimate $E$ by reporting the boundaries
 of a 68.3\% credible region computed with the approximate Beta
 posterior for $E$ given by the method of moments, with inputs $E$ and
 $V$.
 Figure~\ref{fig-betas} shows that finding the approximate Beta
 posterior with the method of moments gives very good results even
 when the input single-sample efficiencies are significantly
 different.  The overall efficiency has been estimated by performing
 10000 repeated pseudo-experiments, and its distribution appears to be
 practically identical to the Beta density found with the method of
 moments.
 In this example, two small ($N_1=26$ and $N_2=10$ events) samples
 from populations with known weights ($w_1=0.7$ and $w_2=0.3$) and
 efficiencies ($\eff_1=0.3$ and $\eff_2=0.7$) have been mixed.  The
 events passing the selection ($K_1=18$ and $K_2=3$) have been used to
 estimate the single-sample efficiencies, whose reference posterior
 densities are shown in the plot with a dashed red line and continuous
 blue line, respectively.  The distribution obtained with the
 pseudo-experiments is best fitted by a Beta density with parameters
 $(25.64,18.97)$ and the approximate method gives a Beta posterior
 with parameters $(25.44, 18.80)$, which is a very good approximation
 (the two curves overlap within their line width).

\subsection{Events with positive or negative unit weights}\label{sec-posneg-weights}

 In high-energy physics simulations, it might happen to work with
 simulated samples generated by MC@NLO \cite{mcatnlo}, which assigns 
 positive and negative (unit) events weights.  Each individual
 event is independently simulated, and knows nothing about its weight.
 Hence we have two categories ($s=2$) with weights $w_1 = +1$ and $w_2
 = -1$, initial sample sizes $N_1=n_{+}, N_2=n_{-}$, and
 $K_1=k_{+}, K_2=k_{-}$ entries after the selection.

 For each sample, the efficiencies $\eff_{+}$ and $\eff_{-}$ can be
 computed individually and inserted into equations
 (\ref{eq-effective-eff}) and  (\ref{eq-effective-var}) above to
 obtain the overall efficiency estimate
\begin{equation}\label{eq-effective-eff-MCNLO}
   E = \frac{\eff_{+} n_{+} - \eff_{-} n_{-}}{n_{+}-n_{-}}
\end{equation}
 with variance
\begin{equation}\label{eq-effective-var-MCNLO}
  V = \frac{n_{+}^2 V(\eff_{+}) + 
         n_{-}^2 V(\eff_{-})}{(n_{+}-n_{-})^2} 
\end{equation}
 where $V(\eff_{+})$ and $V(\eff_{-})$ are the single-sample variances.

 We note that equation~(\ref{eq-effective-eff-MCNLO}) coincides with
 the recommendation by the MC@NLO authors when estimating the
 single-sample efficiency with the success frequency $f_{\pm} =
 k_{\pm}/n_{\pm}$.  In Ref.~\cite{mcatnlo} they say that the
 efficiency should be estimated as $f=(k_{+}-k_{-})/(n_{+}-n_{-})$
 when $k_{+} \ge k_{-}$ or zero otherwise\footnote{It is assumed that
   $n_{+} > n_{-}$ always, because this is a necessary condition for
   the MC@NLO output to be physical.  However, this might be not true
   in the tail of a distribution.  Rebinning might be necessary to
   ensure that this fundamental requirement is satisfied.  Otherwise,
   the only solution is to generate many more events.}.  This ration
 can be rewritten in a form $(f_{+} n_{+} - f_{-}
 n_{-})/(n_{+}-n_{-})$ which gives
 equation~(\ref{eq-effective-eff-MCNLO}) with the replacement
 $f\to\eff$.
 They also suggest to use the usual ``propagation of errors'' to
 estimate its variance whenever the numbers are high enough that the
 Gaussian approximation holds, or to run many MC samples through the
 cuts and look at the dispersion in the result if the data sample is
 too small.  We have obtained equation~(\ref{eq-effective-var-MCNLO})
 with the usual algebra of variances, and in the previous section we
 have seen that it can be used with the method of moments to find an
 approximate uncertainty band which is very close to the result
 obtained with pseudoesperiments (and does not require the Gaussian
 approximation to hold).  It is sufficient to use equations
 (\ref{eq-a-mom}) and (\ref{eq-b-mom}) to find the approximated
 posterior Beta density, which can be used to plot the 68.3\%
 intervals\footnote{The TEfficiency class provided by ROOT now allows
   to plot 68.3\% intervals from any Beta density.}.

 A fully Bayesian treatment of the difference between two independent
 random variables, each one following a Beta distribution, was
 performed by Pham-Gia, Turkkan \& Eng \cite{pham93}, who found an
 analytical solution.  There are few subtleties about their
 ``beta-difference'' distribution , which make it quite complicate to
 be used in the problem considered here:
\begin{itemize}
  \item here we have a weighted difference $e_{+} - e_{-}$ where
    $e_{\pm} = n_{\pm}\eff_{\pm}/(n_{+}-n_{-})$ can be considered 
    ``scaled''  efficiencies; 

  \item the general difference between two Beta-distributed variables
    has domain in $[-1,+1]$ whereas here we require it to be
    restricted to $[0,1]$ to have a physically meaningful quantity.
    This means that the beta-difference distribution has to be
    truncated at zero and renormalized, which requires a numerical
    treatment;

  \item a uniform prior is used (wrongly proposed as non-informative)
    to compute the beta-difference distribution, rather than a
    reference prior.  This is not going to make a dramatic change, as
    shown in section~\ref{sec-ref-post} above, unless the sample size
    is very small.

\end{itemize}
 Given the complexity of their formula and the necessity of accounting
 for the these complications, its seems much easier to proceed in the
 approximate method explained above, which is good (or at least
 acceptable) in most (practically all) common problems.


\subsection{What to do if the samples are not independent?}\label{sec-trigger}

 The case in which the initial dataset does not represent a
 statistically independent sample is expecially important in trigger
 efficiency measurements, when there is no other trigger selection
 which is statistically uncorrelated with respect to the signature $A$
 under study.  Ideally, one would like to select the initial sample
 with a random trigger and then count how many events also survive the
 trigger $A$.  However, for most interesting triggers a randomly
 collected sample contains practically no event which may satisfy
 them, making such random sampling useless.  For this reason, some
 other trigger $B$ is used to select the initial sample, which is
 somewhat correlated with $A$ (for example, $B$ could be a trigger of
 the same type but looser than $A$).  This correlation, which is
 necessary to select the initial sample such that there is some
 fraction of events passing $A$, must be accounted for explicitly when
 computing the efficiency of $A$.

 The trigger efficiency $\eff_{A}$ is the conditional probability that
 a single event, given all other conditions (collider settings,
 detector status and defects, calibration parameters, offline event
 selection, observables $\vec{x}$), is not rejected by $A$.  In
 particular, it is fundamental to notice that different studies
 perform different offline event selections, such that the same
 trigger may have different efficiencies for each of them.  Here we
 assume that the offline event selection is fixed, and we are
 interested into estimating the trigger efficiency $\eff_{A} =
 P(A|\vec{x},\text{off.~sel.},\ldots) \equiv P(A)$ omitting all
 fixed conditions for brevity.  The starting point is the sample of
 events which have been preselected by trigger $B$ and the same
 offline selection (with all other conditions fixed).

 In order to find the desired efficiency $P(A)$, we make use of the
 relation defining the conditional probability, $P(A \cdot B) = P(A|B)
 \, P(B) = P(B|A) \, P(A)$, obtaining
\begin{equation}\label{eq-trig-eff}
   P(A) = P(A|B) \, [P(B) / P(B|A)]  \;,
\end{equation}
 where $P(A|B)$, the conditional trigger efficiency of $A$ for
 events which already passed $B$, can be estimated by taking
 the ratio between the final and initial (i.e.~after $B$) sample
 sizes, as explained in the previous sections.
 
 The fraction in square brackets cannot be determined with real data
 alone.  Even when the trigger efficiency $P(B)$ is measured from real
 data, the conditional probability that the events pass $B$ given that
 $A$ is satisfied cannot clearly be estimated from real data.
 Simulated events are necessary to estimate $P(B|A)$ or the whole
 ratio $P(B) / P(B|A)$, in order to obtain an estimate of the desired
 trigger efficiency $P(A)$. 

 Usually, the auxiliary trigger $B$ is chosen in such a way that one
 can safely assume $P(B|A)=1$ with negligible uncertainty, by virtue
 of what is found with simulated samples.  If this is the case, then
 the impact of the MC events is minimal: one additional measurement
 (performed with an independent sample) is required to find the
 trigger efficiency $P(B)$ of $B$ alone, together with the estimation
 of $P(A|B)$ with the sample preselected by $B$, in order to find
 $\eff_{A}$ from equation~(\ref{eq-trig-eff}).  It is recommended to
 always check that the uncertainty on $P(B|A)=1$ is negligible with
 respect to the uncertainty associated to $P(A|B)$.  If this is not
 the case, one can create an approximate model $P(B|A)$ with a
 monotonically increasing Beta density which peaks at one, and use the
 Bayes' theorem to find the overall probability distribution for the
 parameter $\eff_{A}$ of interest.  Finding an approximate Beta
 density also works when $P(B|A)$ is not 100\%, when the impact of the
 prior information coming from simulated samples is larger.


\section{Summary}

 Estimating the selection efficiency is a fundamental task in most
 data analyses, based on simulated and/or real data.  The measured
 success frequency provides the best estimate of the true efficiency
 in the frequentist approach (being the MLE) and coincides with the
 posterior mode obtained in the Bayesian treatment with the widely
 used uniform prior.  However, such prior cannot be considered
 non-informative.  Instead, if we are completely uncertain about the
 efficiency before making the experiment, or we aim at reporting
 ``objective'' Bayesian results, the use of the reference prior (which
 is the same as the Jeffreys' prior in the binomial model) is
 recommended, together with the intrinsic estimators which have been
 reviewed in section~\ref{sec-intrinsic-estimators}.

 Within the Bayesian approach, if some prior knowledge is available,
 for mathematical simplicity it is recommended to encode it into a
 function belonging to the family of Beta distributions, whose
 parameters can be determined with the method of moments if the exact
 function is not known.  This ensures that the posterior also belongs
 to the same family, so that the math is simplified because all
 properties summarized in appendix~\ref{sec-Beta} are immediately
 available.  An important example of the use of informative priors is
 the combination of independent samples, which is also used for
 including prior knowledge coming from simulations to model systematic
 effects.

 The knowledge of the uncertainty about the efficiency is needed when
 scaling observed quantities to estimate their original values
 (e.g.\ the true rate).  In this case, the easiest approach is to use
 the mean and variance of the posterior density in the computation,
 whenever the use of the full posterior is not practical.  The usual
 variance algebra holds, with the caveat that the square root of the
 final variance might not be good to define a symmetric credible
 interval, because of the inherent asymmetry of the posterior in the
 general case.  Though in many applications the posterior will be
 significantly peaked around the true value, so that the binomial
 (symmetric) approximation holds, care needs to be taken when handling
 very low or very high efficiencies, and when the number of events is
 relatively small, because such approximation behaves very poorly in
 such cases. 

 Several options for the confidence intervals are reviewed in
 \cite{cousins2010}, where it is emphasized that only the
 Clopper-Pearson $q$-confidence intervals never undercover.  At the
 same time, they are considered too wide by many professional
 statisticians, who have proposed a number of recipes for defining
 confidence intervals with the desired average coverage but may
 undercover sometimes.  Among the approximated confidence intervals,
 one may also select the Bayesian posterior $q$-credible intervals,
 which are numerically very similar to widely used frequentist
 approximations but have the advantage of a clear and unanbiguous
 interpretation in terms of the probability that the true unknown
 value is actually contained by them (which is not true for confidence
 intervals).

 When plotting the result of an efficiency measurement, the observed
 frequency should be accompanied by asymmetric error bars.  The
 TEfficiency class of the ROOT framework allows the user to choose
 among a number of options, including Clopper\&Pearson confidence
 intervals, few frequentist approximated confidence intervals, and
 Bayesian credible intervals with different priors.  If the coverage
 is considered an important aspect, then one should choose the
 Clopper\&Pearson intervals.  Otherwise, one may plot the Bayesian
 credible intervals obtained with the reference prior, which are also
 classical confidence intervals very similar to the Wilson
 approximation, achieving the desired coverage on the average (but not
 always).  When the full probability model is not known and only mean
 and variance are available, acceptable intervals may be obtained
 (using TEfficiency) from the approximate Beta density with the same
 mean and variance, for illustration purposes.

 Finally, special care must be used when handling samples that do not
 have unit weights or are not independent.  Few recipes to deal with
 the most common use cases in particle physics have been sketched in
 section~\ref{sec-patologic}, including the case of weighted mixtures
 of MC samples with different cross sections.  Approximate asymmetric
 uncertainty bands can be obtained in a simple way by finding the
 approximate Beta posterior with the method of moments.  Finally, the
 measurement of the trigger efficiency starting from a sample which
 has been preselected by requiring another trigger which is not
 statistically independent is addressed, emphasizing the importance of
 a correct modeling of the relation between the two selections.


\section*{Acknowledgments}

 The author wishes to thank Lorenzo Moneta, who implemented in ROOT
 (starting from v5.28) the TEfficiency class which allows to apply all
 methods discussed in this paper.


\appendix

\section{Useful relations}\label{appendix}

 This appendix summarizes mathematical definitions and properties that
 are useful when dealing with binomial processes.  They can be found
 in standard books like \cite{abramowitz72,BayesianTheory1994}.

\subsection{Gamma function}

 The \emph{Gamma function} is defined on the complex plane ($z \in
 \mathbb{C}$):
\begin{equation}
  \Gamma(z) = \int_0^\infty t^{z-1} e^{-t} \di t
\end{equation}
 with $\Gamma(z+1) = z \, \Gamma(z)$.  For integer values, $\Gamma(n) =
 (n-1)!$.

\subsection{Beta distribution}\label{sec-Beta}

 The Euler \emph{Beta function} is a symmetric function of
 $a,b\in\mathbb{R}$:
\begin{equation}
  B(a,b) \equiv \int_0^1 t^{a-1} (1-t)^{b-1} \di t
         = \frac{\Gamma(a) \, \Gamma(b)}{\Gamma(a+b)}
         = B(b,a)
\end{equation}
 and the \emph{incomplete Beta function} is
\begin{equation}\label{eq-incompl-beta-func}
  B_x(a,b) = \int_0^x t^{a-1}(1-t)^{b-1} \di t \; .
\end{equation}
 with $x \in [0,1]$.

 For $x\in[0,1]$, the \emph{Beta distribution} has probability density
 function 
\begin{equation}\label{eq-Beta-dist}
  f(x;a,b) = \frac{1}{B(a,b)} x^{a-1} (1-x)^{b-1}
           \equiv \Be(x;a,b)
\end{equation}
 and cumulative distribution function 
\begin{equation}
  F(x;a,b) = \int_0^x f(t;a,b) \di t 
           = \frac{B_x(a,b)}{B(a,b)} 
           \equiv I_x(a,b)
\end{equation}
 where $I_x(a,b) = 1 - I_{1-x}(a,b)$ is the \emph{regularized
 incomplete Beta function}.  The mean $E$, mode $m$ (defined only for
 $a>1, b>1$), variance $V$ and skewness $\gamma_1$ of the Beta density
 (\ref{eq-Beta-dist}) are
\begin{eqnarray}\label{eq-Beta-dist-properties}
  E(x;a,b) &=& \frac{a}{a+b} \\
  m(x;a,b) &=& \frac{a-1}{a+b-2} \\
  V(x;a,b) &=& \frac{ab}{(a+b)^2 \, (a+b+1)} \\
  \gamma_1(x;a,b) &=& 
    \frac{2(b-a)\sqrt{a+b+1}}{(a+b+2)\sqrt{ab}}
\end{eqnarray}
 When a single parameter is equal to one the
 density is monotonically decreasing or increasing with a unique
 maximum at $x=0$ or $x=1$; when $a=b=1$ one has the uniform
 distribution.

 Finally, the characteristic function is
\begin{equation}\label{eq-Beta-characteristic-func}
  \phi(t) = \int_0^1 \Be(x;a,b) \exp(-2\pi i x t) \di x 
          = {}_{1}F_{1}(a; a+b; it)
\end{equation}
 where ${}_{1}F_{1}(a;b;c)$ is the confluent hypergeometric function of
 the first kind.

\appendix

\section{Computing intrinsic credible intervals}\label{app-intrinsic-intervals}

 Although it is quite easy to find numerical libraries which minimize
 a function, finding posterior $q$-credible intervals which achieve
 the minimal expected posterior loss is not as simple.  The intervals
 shown in this paper have been computed with a simple numerical
 treatment which, although not being the best possible approach, is
 good enough to provide results which can approximate the true
 intervals with arbitrary precision.  The algorithm explained below
 has been implemented in C++ and executed from within the ROOT
 framework, which offers the user all necessary utilities to perform
 function minimization and compute all special functions which are
 needed in the statistical analysis.

 The simplest approach is to start by computing the values achieved by
 the Beta density which represents our posterior over a grid which
 splits the support $[0,1]$ in equal steps.  Because we wanted to
 report values with 3 decimal places, this interval has been
 subdivided into 2000 steps.  A loop is performed over all points and
 the values of the Beta density and its cumulative distribution
 function are saved into separate arrays, together with the values of
 the loss function.  In addition, the position of the minimum of the
 loss function is saved.

 The second step is to find all intervals which cover a total area of
 $q$.  The search is performed looping over all elements of the
 arrays: a cycle is performed from the first element to the one which
 corresponds to the minimum of the loss function, and a second nested
 cycle is performed from the minimum position to the right edge of the
 support.  All pairs of indices for which the difference in the
 cumulative distribution function equals $q$ within a predefined
 tolerance ($10^{-3}$ in our case) are saved into an STL set.

 The last step is to loop over all pairs of indices.  For each pair,
 the sum of the values which the loss function assumes for all
 intermediate indices is performed.  The pair with the lowest sum is
 the desired interval.




\begin{thebibliography}{99}

\bibitem{pdb08}
C. Amsler et al.,
``Review of Particle Physics'',
Phys. Lett. B667 (2008) 1.

\bibitem{BayesianTheory1994}
 J.M. Bernardo, A.F.M. Smith, \emph{Bayesian theory}, Wiley, 1994

\bibitem{cousins2010}
R.D. Cousins, K.E. Hymes, J. Tucker,
``Frequentist Evaluation of Intervals Estimated for a Binomial
  Parameter and for the Ratio of Poisson Means'', 
NIM A 612 (2010) 388--398;
\arxiv{0905.3831}.

\bibitem{vandenbos2007}
A. Van den Bos,
\emph{Parameter Estimation for Scientists and Engineers},
New York, Wiley, 2007. 

\bibitem{james06}
F. James,
``Statistical Methods in Experimental Physics: 2nd Edition'',
World Scientific, 2006.

\bibitem{brown2001}
 L.D. Brown, T.T: Cai, A. DasGupta,
 ``Interval Estimation for a Binomial Proportion'',
 Stat.Sci. 16 (2001) 101-133

\bibitem{root-cpc}
I. Antcheva et al.,
``ROOT --- A C++ Framework for Petabyte Data Storage, Statistical
Analysis and Visualization'',
Computer Physics Communications, Vol. 180, Issue 12 (2009) 2499--2512
\doi{10.1016/j.cpc.2009.08.005}
(ROOT web site: \url{http://root.cern.ch}).

\bibitem{cousins0702156}
  R.D.~Cousins, J.T.~Linnemann, \& J.~Tucker,
  ``Evaluation of three methods for calculating statistical
  significance when incorporating a systematic uncertainty into a test
  of the background-only hypothesis for a Poisson process'',
  NIM A 595 (2008) 480--501,
  \doi{10.1016/j.nima.2008.07.086},
  \arxiv{physics/0702156}.

\bibitem{casadei2012}
D. Casadei,
``Reference analysis of the signal + background model in counting
experiments'',
JINST 7 (2012) 01012,
\doi{10.1088/1748-0221/7/01/P01012}, \arxiv{1108.4270}.

\bibitem{bernardo05a}
J.M. Bernardo,
``Reference analysis'',
Handbook of Statistics 25 (D.K. Dey and C.R. Rao eds.). Amsterdam:
Elsevier (2005) 17--90.

\bibitem{berger09}
J.O. Berger, J.M. Bernardo, D. Sun,
``The formal definition of reference priors'',
Annals of Statistics 37 (2009) 905--938.

\bibitem{bernardo2007}
J.M. Bernardo,
``Ojective Bayesian point and region estimation in location-scale models'',
Sort 31 (2007) 3-44.

\bibitem{jeffreys46}
H. Jeffreys,
``An invariant form for the prior probability in estimation
problems'',
Proc. Royal Soc. London A Math. and Phys. Sci., vol. 186, no. 1007
(1946) 453--461.

\bibitem{Bernardo2003}
 J.M. Bernardo, ``Bayesian Statistics'', 
       \emph{Encyclopedia of Life Support Systems (EOLSS). Probability
       and Statistics} (R. Viertl, ed.),
       Oxford, 2003, UNESCO (\url{http://www.eolss.net}).



\bibitem{bernardo05b}
J.M. Bernardo,
``Intrinsic credible regions: An objective Bayesian approach to interval
estimation'',
Test 14 (2005) 317--384.

\bibitem{mcatnlo}
S. Frixione and B.R. Webber,
``Matching NLO QCD computations and parton shower simulations'',
JHEP 06 (2002) 029; \arxiv{hep-ph/0204244}.

\bibitem{pham93}
T. Pham-Gia, N. Turkkan, P. Eng,
``Bayesian analysis of the difference of two proportions'',
Comm. Statist.---Theory Meth., 22:6 (1993) 1755--1771.

\bibitem{abramowitz72}
Abramowitz, M. and Stegun, I. A. (Eds.),
``Handbook of Mathematical Functions with Formulas, Graphs, and
Mathematical Tables'', 
10-th printing. New York: Dover, 1972.

\end{thebibliography}
\end{document}